\documentclass[12pt,preprint]{aastex}

\usepackage{graphicx}
\usepackage{natbib}

\shorttitle{Chandra Observations of low $\sigma$ groups}
\shortauthors{Helsdon, Ponman \& Mulchaey}
 
\begin{document}




\bibliographystyle{apj}


\title{Chandra Observations of low velocity dispersion groups}


\author{Stephen F. Helsdon\altaffilmark{1}}
\affil{The Observatories of the Carnegie Institute of Washington, 813 Santa Barbara Street, Pasadena, CA 91101, USA\\
and\\
School of Physics and Astronomy, University of Birmingham, Edgbaston, Birmingham B15 2TT, UK}

\author{Trevor J. Ponman}
\affil{School of Physics and Astronomy, University of Birmingham, Edgbaston, Birmingham B15 2TT, UK}

\and

\author{J. S. Mulchaey}
\affil{The Observatories of the Carnegie Institute of Washington, 813 Santa Barbara Street, Pasadena, CA 91101, USA}

\altaffiltext{1}{sfh@star.sr.bham.ac.uk}


\begin{abstract}
  Deviations of galaxy groups from cluster scaling relations can be
  understood in terms of an excess of entropy in groups.  The main effect
  of this excess is to reduce the density and thus luminosity of the
  intragroup gas. Given this, groups should also should show a steep
  relationship between X-ray luminosity and velocity dispersion.  However,
  previous work suggests that this is not the case,
  with many measuring slopes flatter than the cluster relation.\\
  Examining the group $L_X:\sigma$ relation shows that much of the
  flattening is caused by a small subset of groups which show
  very high X-ray luminosities for their velocity dispersions (or vice versa).\\
  Detailed Chandra study of two such groups shows that earlier ROSAT
  results were subject to significant ($\sim$30-40\%) point source
  contamination, but confirm that a significant hot IGM is present in these
  groups, although these are two of the coolest systems in which
  intergalactic X-ray emission has been detected.\\
  Their X-ray properties are shown to be broadly consistent with those of
  other galaxy groups, although the gas entropy in NGC\,1587 is unusually
  low, and its X-ray luminosity correspondingly high for its temperature,
  compared to most groups.\\
  This leads us to suggest that the velocity dispersion in these systems
  has been reduced in some way, and we consider how this might have come
  about.
\end{abstract}




\section{Introduction}
\label{sec:intro}

Detailed X-ray studies of the hot intergalactic medium (IGM) in groups
became possible for the first time with ROSAT \citep{mulchaey96,ponman96},
allowing the basic properties of the IGM in groups to be established and
compared to that in rich clusters. Groups are found to depart
systematically from cluster trends, showing lower X-ray luminosity at a
given temperature than would be expected from extrapolation of the cluster
$L:T$ relation \citep{ponman96,helsdon00} and flatter X-ray surface
brightness profiles than clusters \citep{ponman99}. It has proved
instructive to consider these observations in terms of the {\it entropy} of
the gas \citep{bower97,ponman99}, which is found to be higher in low mass
systems than would be expected from a self-similar scaling of clusters,
although the idea of a simple universal ``entropy floor'' \citep{ponman99}
now looks to have been over-simplistic
\citep{pratt03,ponman03,sun03,mushotzky03,voit03}.

A number of mechanisms have been suggested which might account for this
behavior of the entropy, by heating of the gas, or removing low entropy gas
through cooling, or some combination of the two -- a review of these models
can be found in \citet{ponman03}. Irrespective of the means by which the
entropy is raised in lower mass systems, the main effect is to reduce the
density of the IGM, though its temperature may also be somewhat increased
(e.g. \citealt{voit02}). Hence, if this picture for the similarity breaking
is basically correct, one would also expect the $L_X$:$\sigma$ relation
(between X-ray luminosity and velocity dispersion) to steepen in galaxy
groups, since the reduction in gas density lowers $L_X$ but not $\sigma$.
However, a variety of studies (e.g.
\citealt{mulchaey98,helsdon00b,mahdavi00}) have shown that this is not the
case, although more recently \citet{mahdavi01} has argued that there is a
continuous, and gradually steepening $L_X$:$\sigma$ relation from clusters
to individual galaxies. \citet{helsdon00b} assembled a sample of 42 galaxy
groups, and derived a slope for the $L_X$:$\sigma$ relation of between
2.4$\pm$0.4 and 4.7$\pm$0.9 (depending on how the data were fit). This
group slope is flatter (or at least not steeper) than the cluster relation,
which has been variously derived as 6.38$\pm$0.46 \citep{white97},
5.24$\pm$0.29 \citep{wu99} and 4.4$^{+0.7}_{-0.3}$ \citep{mahdavi01} -- all
these results being obtained using orthogonal regression techniques.

A flat $L$:$\sigma$ relation in groups is also indicated by the independent
studies of \citet{mulchaey98} and \citet{mahdavi00}. The former authors had
only a small sample of groups, but derived reliable dispersions for these
by enlarging their group membership to $\ga$20 galaxies via multi-fiber
spectroscopy. Mahdavi et al, used ROSAT All Sky Survey data, allowing the
study of a large sample of groups, but providing poor quality data (a few
hundred seconds exposure) for each one. Hence they were unable to remove
X-ray emission arising from individual galaxies within the systems they
studied. They derived an $L$:$\sigma$ relation which flattens drastically
to $L_X \propto \sigma^{0.37}$ in the group regime.  Unfortunately, it is
unclear to what extent this results from contamination of their diffuse
emission with flux from individual group galaxies.

Thus, these group studies suggest that the $L_X:\sigma$ relation in groups
is actually \textit{flatter} than an extrapolation of the relation for
clusters. This surprising result could indicate a possible fundamental flaw
in the whole picture whereby the X-ray luminosity in poor systems is
suppressed by excess entropy. Alternatively, it might result from the
existence of unexpected new sources of X-ray emission in some low velocity
dispersion groups. Looking at the group $L_X:\sigma$ relation (e.g. Figure
4 in \citealt{helsdon00b}) it is clear that the flat $L_X:\sigma$ relation
is driven by the properties of a set of groups with $\sigma\la$110 km
s$^{-1}$ which have diffuse luminosities much greater than the values which
would be expected on the basis of the cluster $L$:$\sigma$ relation ($L_X
\ll 10^{41}$ erg s$^{-1}$).

In order to better understand the properties of these low velocity
dispersion groups we have obtained Chandra data for two of these systems.
In section~\ref{sec:reduction} we describe the spatial and spectral
analysis, and present some basic properties of these systems. In
section~\ref{sec:roscomp} we compare the Chandra and ROSAT observations of
these systems, and finally we discuss the results in section~\ref{sec:dis}.
Throughput this paper we use H$_0 = 75$ km s$^{-1}$ Mpc$^{-1}$.

\section{Data Reduction and Analysis}
\label{sec:reduction}

We obtained observations of two low velocity dispersion groups, a $\sim$20
ksec observation of NGC\,1587 on October 3rd 2000, and a $\sim$20 ksec
observation of NGC\,3665 on 3rd November 2001. One point of concern with
these two systems is the fact that their velocity dispersions are so low
(106 km s$^{-1}$ for NGC\,1587 and 29 km s$^{-1}$ for NGC\,3665). In their
original group catalogues these systems have had their velocity dispersions
estimated from a small number of galaxies. A natural concern is therefore
that the anomalous positions of these systems in the $L_X:\sigma$ plot
might arise from seriously inaccurate values of $\sigma$. However,
searching the NED database for additional galaxies lying within 1 Mpc in
projection, and 400~km~s$^{-1}$ in velocity from the cataloged group
centroid, brings the membership to 8 for NGC\,1587 and 10 for NGC\,3665.
While these extra members do tend to increase the calculated dispersions of
the groups a little, both these systems still have low velocity dispersions
-- 108 km s$^{-1}$ for NGC\,1587 and 65 km s$^{-1}$ for NGC\,3665.

The basic X-ray data reduction was based on the ``CIAO science threads''
given on the Chandra X-ray Center (CXC) web pages, and is briefly
summarised below. All analysis was carried out using CIAO version 2.3, and
CALDB version 2.21.

\subsection{Initial Reduction}

Initially the data were reprocessed to ensure the latest calibration files
were used (e.g. gain), and the CTI correction applied. In addition, the
$\sim$1.5 arc second aspect offset present in some Chandra observation
files was corrected. The observation of NGC\,3665 was taken in VFAINT mode
(NGC\,1587 was in FAINT mode), and as a result the VF option was selected
for the reprocessing of this dataset as this can significantly reduce the
background at low and high energies in VFAINT observations. Following this,
the wavdetect tool was used to source search a coarsely binned image and
identify obvious sources. These sources were removed from the data and a
light-curve was generated from the remaining data. This light-curve was
then used to identify and remove periods of high background from the data.

The method used to identify periods of high background depends partly on
how subsequent reduction/analysis is to be carried out. In this case, we
make extensive use of the blank sky datasets available in the CALDB, and so
must clean the data in a manner comparable with the method used for the
blank sky files. In this case we actually generate light-curves in two
different energy bands --- for the FI chips we use a energy band of 0.3-12
keV and for the BI chips we use a band of 2.5-7 keV. These energy bands
were selected as these are the bands which best show flares on the two
types of chips (see discussion of the ACIS background on the CXC web
pages). Light-curve bin lengths of 259.28 seconds were used and a 2.5 sigma
clip was applied to remove bad regions of the observation. Note that this
is different to the fixed factor of 1.2 used in the analysis thread. The
2.5 sigma clip was used as the expected quiescent scatter of the 2.5-7 keV
band for the whole S3 chip is greater than a factor of 1.2. Furthermore, if
large regions of the FI chips are excluded (due to sources for example)
then the expected scatter from the FI chips can also exceed the factor of
1.2. In cases like this the blank sky ``cookbook'' recommends extending the
bin length so that the expected scatter is less than a factor of 1.2.
Instead we have used a cut based on the observed scatter which should have
a similar effect.

After identifying times of bad data, the data were filtered on both the
good time intervals inferred from the BI and FI chips. This was done so as
to be as careful as possible to exclude possible flares from the data. For
NGC\,1587 the final exposure time was $\sim$19 ksec and for NGC\,3665
$\sim$17 ksec. Finally, the appropriate blank sky files for the
observations were identified. It was checked that the same calibration
files (e.g. gain) were used for the background files as for the data, and
the background datasets were then reprojected to have the same pointing as
the groups observations.

\begin{figure}
\centering
\includegraphics[totalheight=6.7cm,angle=0]{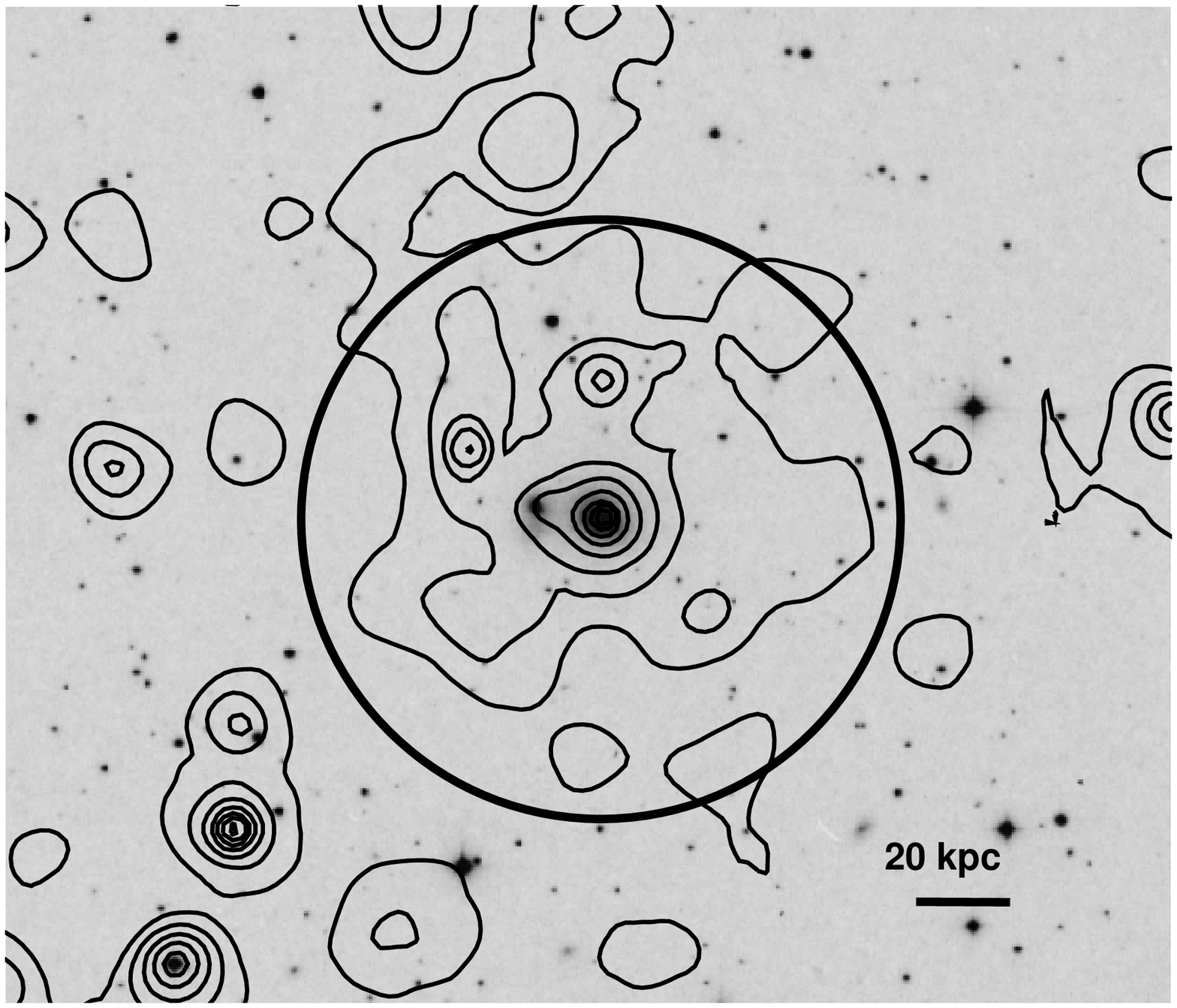}
\includegraphics[totalheight=6.7cm,angle=0]{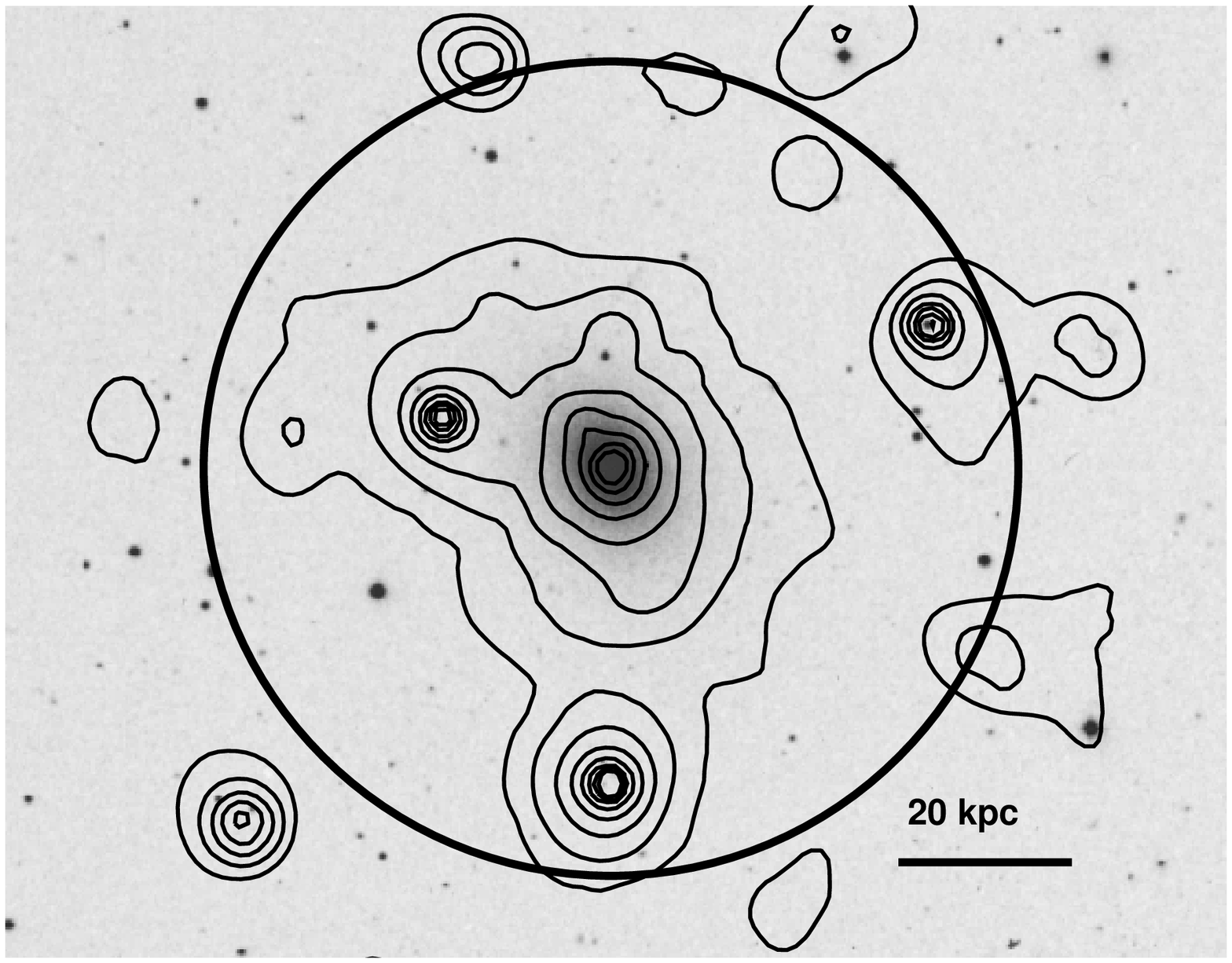}
\caption{\label{fig:images}Adaptively smoothed X-ray contours in the 
  0.7 - 1.5 keV band overlaid onto DSS images, for NGC\,1587 (left) and
  NGC\,3665 (right). The lowest contour is 5 sigma above the background and
  the remaining contours are logarithmically spaced beginning with 10
  sigma, 20 sigma, 40 sigma, etc. The circles show the radii used for
  analysis of the systems.}
\end{figure}

In Figure~\ref{fig:images} adaptively smoothed X-ray contours (in the 0.7 -
1.5 keV band) are overlaid onto a digital sky survey image for both of the
systems. The smoothing was carried out with a gaussian filter whose sigma
was adjusted to be 2/3 the radius needed to include a minimum of 25 counts
in a circle around each pixel. After examining these images and also 1D
profiles of the ratio of data to background, circular regions were selected
in which to carry out further data analysis (e.g. spectral and spatial
analysis). These regions are also marked in Figure~\ref{fig:images}. For
NGC\,1587 the region of interest is a circle of radius 4.5 arcmin and for
NGC\,3665, a circle of radius 5.8 arcmin.

The blank sky datasets were also checked to see if they were a good
representation of the real background in the target observations. The X-ray
contour maps were used to identify several regions where there were
apparently no sources. In each of these regions the observed counts in
several different energy bands were compared with the predicted counts from
the blank sky datasets in the same region. These checks showed that the
predicted background was indeed consistent with that observed.

\subsection{Spectral Analysis}

As we are primarily interested in the diffuse emission from these targets
it is important to remove any contaminating sources. The wavdetect tool in
CIAO was used to source search a full resolution image. The identified
sources were examined in ds9 to ensure all obvious contaminating sources
were identified and that a significant fraction of the diffuse emission was
not removed (e.g. check that a large part of the group central region was
not removed as a ``contaminating source''). All events associated with
these sources were then removed from the data along with all data outside
the region of interest (described in the previous section).

Weighted spectral responses for the data were then generated. The 0.5-2keV
band was selected for the weight map as this produces a map that should
more closely follow the distribution of source counts (most source counts
fall in this band and a harder map would tend towards an area weighting).
Unfortunately, it turns out that in the detection radius for these two
groups, the background still dominates over the source counts. It is not
possible to do a simple background subtraction to recover the source counts
distribution, as the average background counts per pixel is low ($\sim$
0.03 counts per pixel in the blank sky files), and subtracting this number
from the detected source counts (which have a minimum value of 1) will not
have much effect on the distribution (given that negative weights are
correctly ignored).

In order to recover the approximate background subtracted spatial source
distribution, an image of the data and background (normalized to the same
exposure time as the data) are adaptively smoothed with a top hat filter,
which varies in size to include 10 counts in total. This smooth also
records for each pixel the area needed to include these 10 counts. For each
pixel, by comparing the area needed in the background image with the area
needed in the source image, it is possible to obtain an estimate of the
ratio of source flux to background flux. Given this ratio, and an estimate
of the background, it is possible to recover an estimate of the source
distribution. This estimate of the source distribution was then used to
generate the weighted responses. The resultant effective area file was also
corrected for the time dependent reduction in the ACIS low energy quantum
efficiency using the tool, corrarf, available on the CXC web pages.

\begin{figure}
\centering
\includegraphics[totalheight=10cm,angle=0]{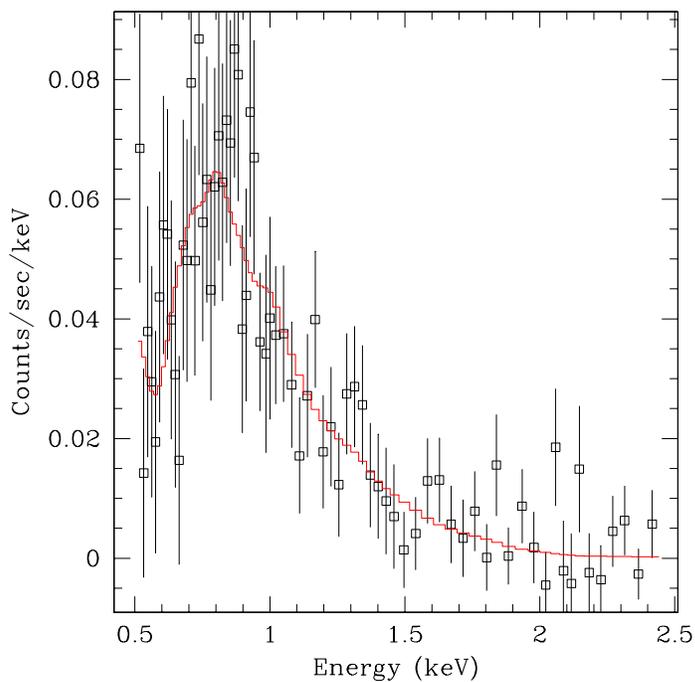}
\caption{\label{fig:spec_fit_ex}Example spectral fit for NGC1587 with 
the Abundance free.}
\end{figure}

Source and blank sky spectra were extracted in the region of interest. The
source spectra were grouped to give a minimum of 20 counts in each bin (to
allow $\chi^2$ fitting later). The background subtracted data were then fit
to a MEKAL plasma model in sherpa \citep{freeman01}.
The fit was restricted to the range
0.5-2.5keV, as there may be calibration problems below 0.5 keV and the
source counts had dropped to essentially zero by 2.5 keV in these systems.
The Hydrogen column density in these fits was fixed at a value determined
from radio surveys \citep{dickey90}. An example spectrum and fit, is shown
in Figure~\ref{fig:spec_fit_ex}. Overall, there were approximately 750 net
source counts for NGC 1587 and approximately 650 for NGC 3665.

After fitting the data, the best fit model was used to generate a
spectral weights file, which contains the fraction of the incident
flux falling in a series of energy bands. This can then be used to
refine the spectral fit --- the weighted spectral responses generated
initially are based on using the detected counts to weight the
responses, whereas, in reality, the incident flux should be used. A
spectral weights file can be used to correct the detected counts to
give the incident flux, which can then be used to generate the
weighted responses (Note that this is discussed in the CXC analysis 
thread on weighting ARFs and RMFs). Updated spectral response files were then
generated and the spectral data were refit. This iterative process was
then repeated until the best fit spectral model stabilized (less than
a 1\% change in best fit temperature and also the lower bound
temperature error was the stable criteria). For both the groups in
this work a stable solution was reached after just two iterations. The
best fitting models for the two groups are given in
Table~\ref{tab:specfits} along with the inferred luminosities. The
luminosities quoted are to the radius of data extraction and include a
small ($<$2\%) correction for diffuse flux which was removed when
excluding other sources. This correction was calculated by comparing
spatial models (described below) with and without sources
removed. Errors on the luminosities are derived using a Monte Carlo
method. A series of spectral models was generated, with a range of
temperatures and abundances, extracted at random
from a gaussian distribution centered on the best fit value, and with
standard deviation corresponding to
the 1$\sigma$ errors. Each model was then refit to determine
the normalisation and a corresponding flux. The standard
deviation of the fluxes was then used to infer the error on the
luminosity.

\begin{deluxetable}{lcccccc}
\tabletypesize{\small}
\tablecaption{\label{tab:specfits}Results of Spectral 
fitting to group emission. }
\tablewidth{0pt}
\tablehead{
\colhead{Group}    & \colhead{N$_H$}              &  \colhead{Temperature}          & \colhead{Abundance}          & \colhead{$\chi^{2}$/dof} & \colhead{Distance} & \colhead{log L$_X$} \\
\colhead{~}        & \colhead{10$^{22}$cm$^{-2}$} &  \colhead{(keV)}                & \colhead{(Solar)}            &                          & \colhead{(Mpc)}    & \colhead{(erg s$^{-1}$)}
}

\startdata
NGC 1587 & 0.0685              &  0.37$^{+0.04}_{-0.04}$ & 0.012$^{+0.012}_{-0.009}$  & 49.76/69       & 48.8           & 41.22 $^{+0.05}_{-0.04}$ \\
NGC 1587 & 0.0685              &  0.53$^{+0.07}_{-0.04}$ & (0.3)                      & 72.76/70       & 48.8           & 40.88 $^{+0.01}_{-0.01}$ \\
NGC 3665 & 0.0206              &  0.45$^{+0.07}_{-0.04}$ & 0.04$^{+0.02}_{-0.01}$     & 92.84/71       & 27.8           & 40.58 $^{+0.05}_{-0.04}$ \\
NGC 3665 & 0.0206              &  0.52$^{+0.06}_{-0.04}$ & (0.3)                      & 106.18/72      & 27.8           & 40.37 $^{+0.01}_{-0.01}$ \\
\enddata 
\tablecomments{All errors are 1$\sigma$. Luminosities are absorption
  corrected, bolometric and assume H$_0=$ 75 km s$^{-1}$.}
\end{deluxetable}

\begin{figure}
\centering
\includegraphics[totalheight=16cm,angle=270]{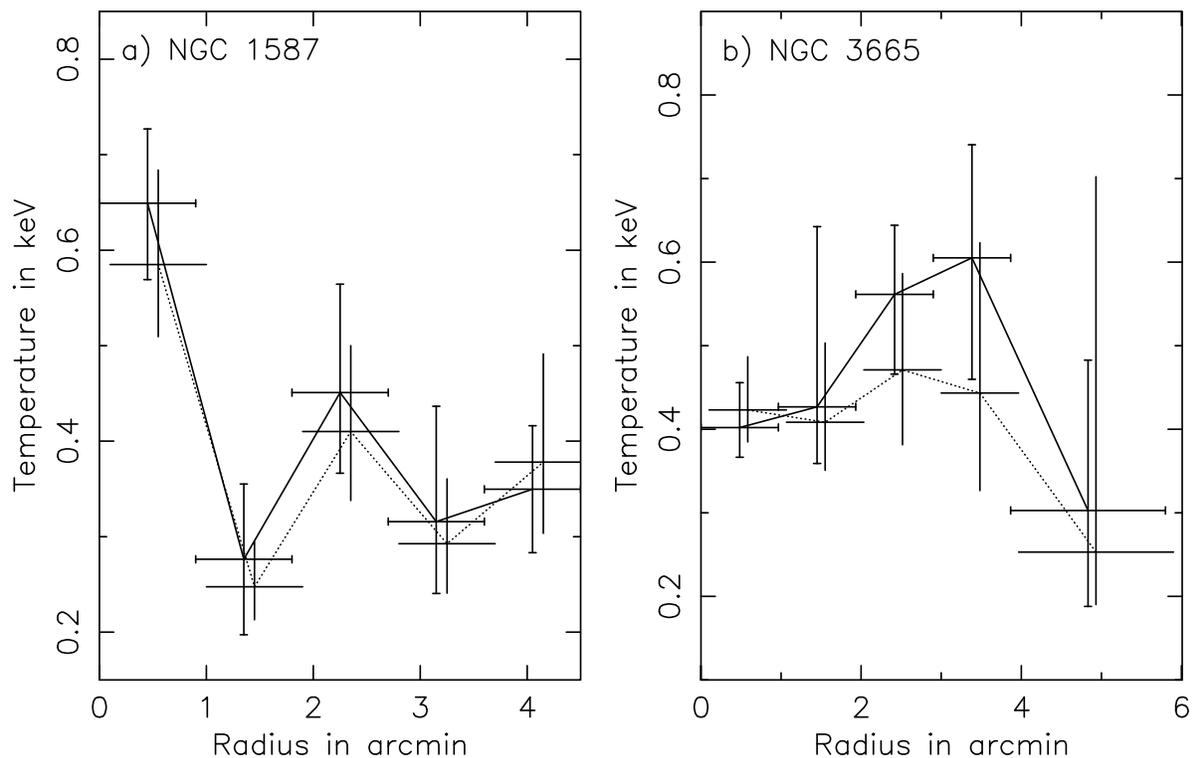}
\caption{\label{fig:2d_tprof} 2D temperature profiles for both a) NGC\,
  1587 and b) NGC3665. Barred crosses and solid line represent the profile
  obtained with the metallicity allowed to vary for each annulus, while the
  plain crosses and dotted line mark the profiles obtained with the
  metallicity fixed at the global value. Note that the fixed metallicity
  points have been offset slightly on the radius axis so as to show the
  error bars more clearly.}
\end{figure}

We were also interested in looking for potential radial temperature
and/or metallicity variations in these systems. As a result, spectra
were extracted in a series of annuli, $\sim$1 arcmin in width, for
each system. We then used the procedure described above for the total
group emission to fit the spectral data in each annular bin. The only
difference from the earlier process was to use the spectral weights
file from the integrated group fit for each of the annular bins,
rather than running the full iterative process on each annular bin. A
test case was run on one of the annuli which confirmed that this did
not significantly alter the results. The projected 2D temperature
profiles are shown in Figure~\ref{fig:2d_tprof}. The abundance
profiles are not shown, as outside of the central bin the abundance was
poorly constrained. For the remainder of this paper, unless explicitly 
stated otherwise, we use an abundance fixed at the global value. Fixing
the abundance at the global value does not have a significant impact
on the derived 2D temperature profiles, as can be seen in 
Figure~\ref{fig:2d_tprof}.

Finally, we also carried out a 3D deprojection of the spectral data in each
annulus. In reality the spectral data for each of the 2D annuli has some
contribution from gas which lies at a larger 3D radius, but which is
projected onto a smaller radius when viewed. We attempt to correct for the
effects of this, by deprojecting the data using an ``onion skin'' method.
Starting with the outer shell (which in theory should have no other
emission projected onto it), and assuming spherical symmetry, it is
possible to calculate what fraction of the volume of the 3D shell is seen,
and to calculate what fraction of this outer shell is projected onto all
inner annuli. It is then possible to construct a series of spectral models
which describe the contribution from this outer 3D shell to each 2D
annulus. In each model, all parameters apart from the normalisation are
fixed at the same values and the volume factors can be used to fix the
relative normalisations of the models. This procedure can then be repeated
for all inner annuli. Thus the spectral data for any particular 2D annulus
is comprised of contributions from the model describing the 3D shell at
this radius, plus contributions from all overlying shells.

\begin{figure}
\centering
\includegraphics[totalheight=16cm,angle=270]{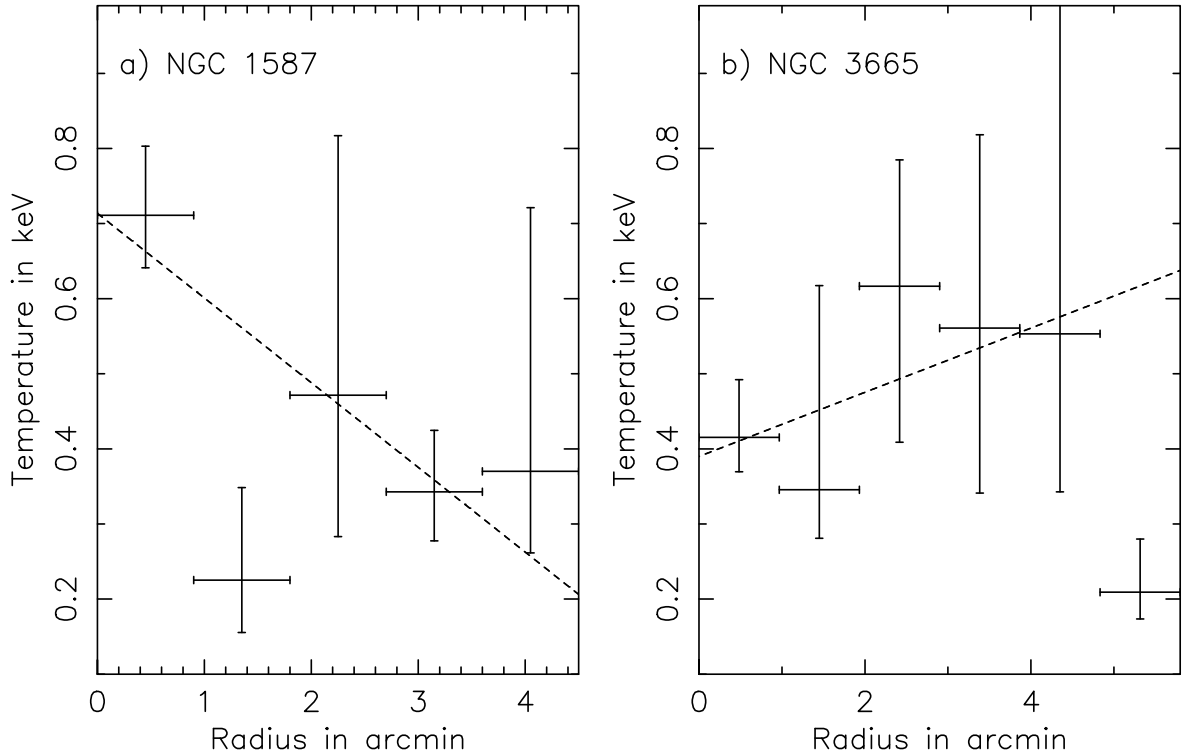}
\caption{\label{fig:3d_tprof} Deprojected 3D temperature profiles for 
  both a) NGC\,1587 and b) NGC3665. The dashed line represents the profile
  obtained if the temperature is constrained to be a linear function with
  radius during fitting.}
\end{figure}

All spectra (and appropriate responses) for each annulus were then
simultaneously fit to the described set of spectral models. This gave a set
of models describing the 3D spectral properties of the gas. As well as
allowing the temperature and metallicity to vary separately in each shell,
models were fit which constrained both the temperature and metallicity to a
simple linear profile with radius. The advantage of using linear models for
the temperature and abundance is that it enables a better impression of
rough radial trends in low count data. This is because only 2 parameters
are fitted (central value and gradient) rather than N parameters, where N
is the number of radial annuli. The deprojected temperature profiles are
shown in Figure~\ref{fig:3d_tprof} and the parameters for the linear
temperature and abundance fits are given in Table~\ref{tab:deproj_lin}.
Note that these radial profiles only cover the inner regions of these
groups, and that it is not sensible to extrapolate them substantially
(for example, to the Virial radius), as the extrapolated profiles may
become unphysical (e.g. negative temperature). It is also worth noting
that by looking at Figure~\ref{fig:3d_tprof} one can see that neither
group is significantly non-isothermal apart from the central point in
NGC 1587.

\begin{deluxetable}{lcccc}
\tabletypesize{\small}
\center{\tablecaption{\label{tab:deproj_lin}Results of the deprojection 
analysis, with both the Temperature ($T$) and abundance ($Z$) profiles
constrained to be simple linear models.}}
\tablewidth{0pt}
\tablehead{
\colhead{Group}   & \colhead{$T_0$}           & \colhead{d$T$/d$r$}               & \colhead{$Z_0$}           & \colhead{d$Z$/d$r$}         \\
                  &  \colhead{(keV)}          &  \colhead{(keV/arcmin)}           &  \colhead{(Solar units)}  & \colhead{(Solar units/arcmin)}       }
\startdata
NGC~1587          & 0.712$^{+0.073}_{-0.086}$ & -0.112$^{+0.035}_{-0.023}$        & 0.118$^{+0.072}_{-0.049}$ & -0.031$^{+0.018}_{-0.018}$ \\
NGC~3665          & 0.390$^{+0.071}_{-0.052}$ & 0.043$^{+0.030}_{-0.031}$         & 0.000$^{+0.019}_{-0.059}$ & 0.042$^{+0.023}_{-0.017}$     \\
\enddata
\tablecomments{Negative values for the gradients indicate a profile dropping with increasing radius.}
\end{deluxetable}

\subsection{Spatial Analysis}

As for the spectral analysis, we are interested in the properties of the
diffuse gas, so all other contaminating sources were removed, along with
all data outside our main region of interest. An image was then generated
in the 0.7-1.5keV band. This particular band was chosen to help maximise
the signal to noise ratio of the data. As well as this image, two other
important sets of information are needed: an exposure map and a background
model. The exposure map takes into account potential exposure variations
across the region of interest.  As a result an exposure map was generated,
using the standard CIAO tools and the spectral weights file derived from
the best fitting integrated group spectral model. We also removed (set to
zero) regions of the exposure map where sources had been removed in the
group image.

Ideally, if the sky background is flat, the detector background should have
the same shape as the exposure map. However comparison of the blank sky
datasets in the 0.7-1.5 keV band with the exposure maps, shows that there
are gradients in the background level across a chip, which are greater than
would be expected from the exposure map alone (by a factor of $\sim$5 in
the worst case). These variations are most likely due to variations in the
cosmic ray component of the background (e.g. see the discussion of the ACIS
background on the CXC web pages). It is important to allow for these
variations, as the diffuse group emission may be at a level close to that
of the background. To do this, a carefully smoothed version of the blank
sky image was used to correct an exposure map image for the effects of this
gradient. The reason a corrected exposure map was used as the basis of the
background, rather than just the smoothed blank sky images, relates to the
procedure used to infer the shape of the background. To produce a smoothly
varying background from the blank sky files, they first had to be binned up
to a much coarser resolution, after which two different smoothing
algorithms were used to infer the structure of the background over a chip.
This smoothing has the side-effect of effectively adding in extra counts
outside the boundaries of the chips, which are not wanted. Using this
smoothed image to correct an exposure map automatically removes these extra
counts (because the exposure map is zero at those points).

This gave a reasonably smooth background image, with an overall shape the
same as that observed in the blank sky dataset. The real blank sky dataset
was then used to predict the overall background for the group observation,
and the background model was renormalised to match this prediction. This
gave a background image with the correct shape and normalisation for the
observation. Finally, this background image was divided by the exposure map
to produce an exposure corrected background model, which would be used as a
model component in Sherpa.

When modelling the diffuse emission, all fits are done with 2D data. Although
1D profile fits would be simpler to derive, they suffer from potentially
serious biases relating to where the profile is centered, and the effects of
possible ellipticity
of the emission \citep{helsdon00}. To speed up the 2D fitting process,
images, exposure maps and background models were actually generated for three
different bin sizes covering
different regions - for data within 1 arcmin of the center a bin size of
$\sim$2 arcsec was used, for data between the radii of 1 and 3 arcmin a bin
size of $\sim$10 arcsec was used, and for radii beyond 3 arcmin a bin size
of $\sim$30 arcsec was used. Thus at larger radii (and lower counts) a
larger bin size was used. This prevented the fitting process from being too
slow, whilst still allowing good resolution in the central regions, which
was important for getting good constraints on the fitted models. The data
were then input to Sherpa and fit in 2D with $\beta$ models of the form:
\begin{equation}
\Sigma(r)=\Sigma_0(1+(r/r_{core})^{2})^{-3\beta_{fit}+0.5} + B
\end{equation}
where the free parameters were the central surface brightness $\Sigma_0$,
the core radius $r_{core}$, the index $\beta_{fit}$ and the $x$ and $y$
position of the center of the emission. Note that models for the different
resolutions were linked together as appropriate. Both spherical and
elliptical fits were carried out on the data, with the ellipticity and the
position angle being extra free parameters in the elliptical fits. The
background component, $B$ was also present in all fits, and it was defined
by a gridmodel component derived from the exposure corrected background
model image generated earlier. Note that this background component was
not free to vary. Given that the cosmic ray background can vary from
one observation to another, test fits in which the background amplitude
was allowed to vary, showed that while both the core radius and beta 
increase (by about 10\% for beta), the new values and associated errors were 
consistent, at approximately the one sigma level, with the original values 
in which the background was frozen. 

Many groups are observed to have 2
components present in their X-ray emission (e.g.
\citealt{mulchaey98,helsdon00}), and we attempted to fit both one and two
component $\beta$ models to the data. Unfortunately, the data for these two
systems were not sufficient to provide robust constraints on any reasonable
two component fits, so it was only possible to derive reliable one
component fits. The results of the 2D fitting are shown in Table
\ref{tab:spatfits1}. These results are for spherical $\beta$ models.
Elliptical $\beta$ models were also fit to the data, but in both cases it
was found that the fit statistic did not improve significantly for the
extra parameters.

\begin{figure}
\centering
\includegraphics[totalheight=8cm,angle=0]{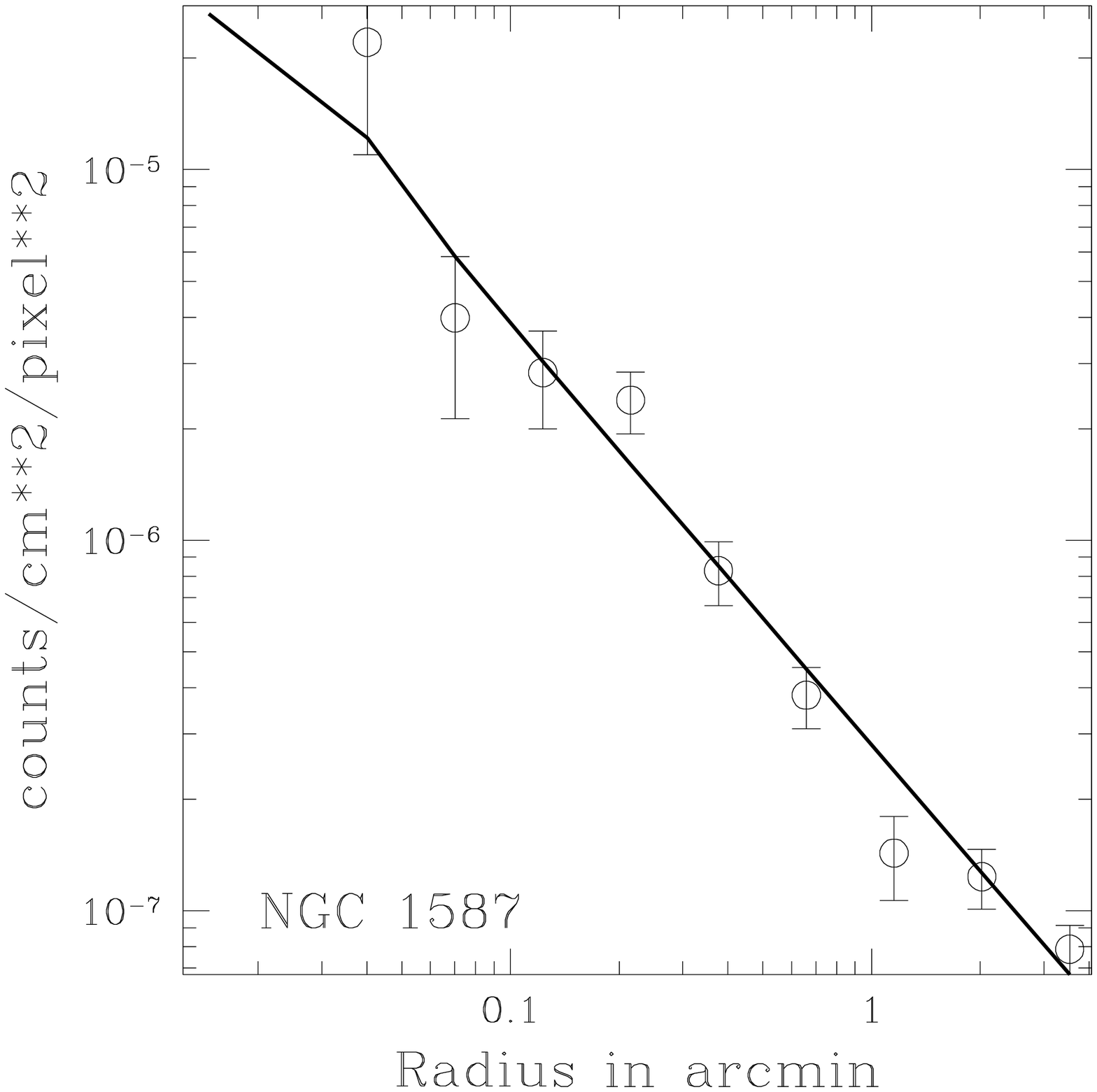}
\includegraphics[totalheight=8cm,angle=0]{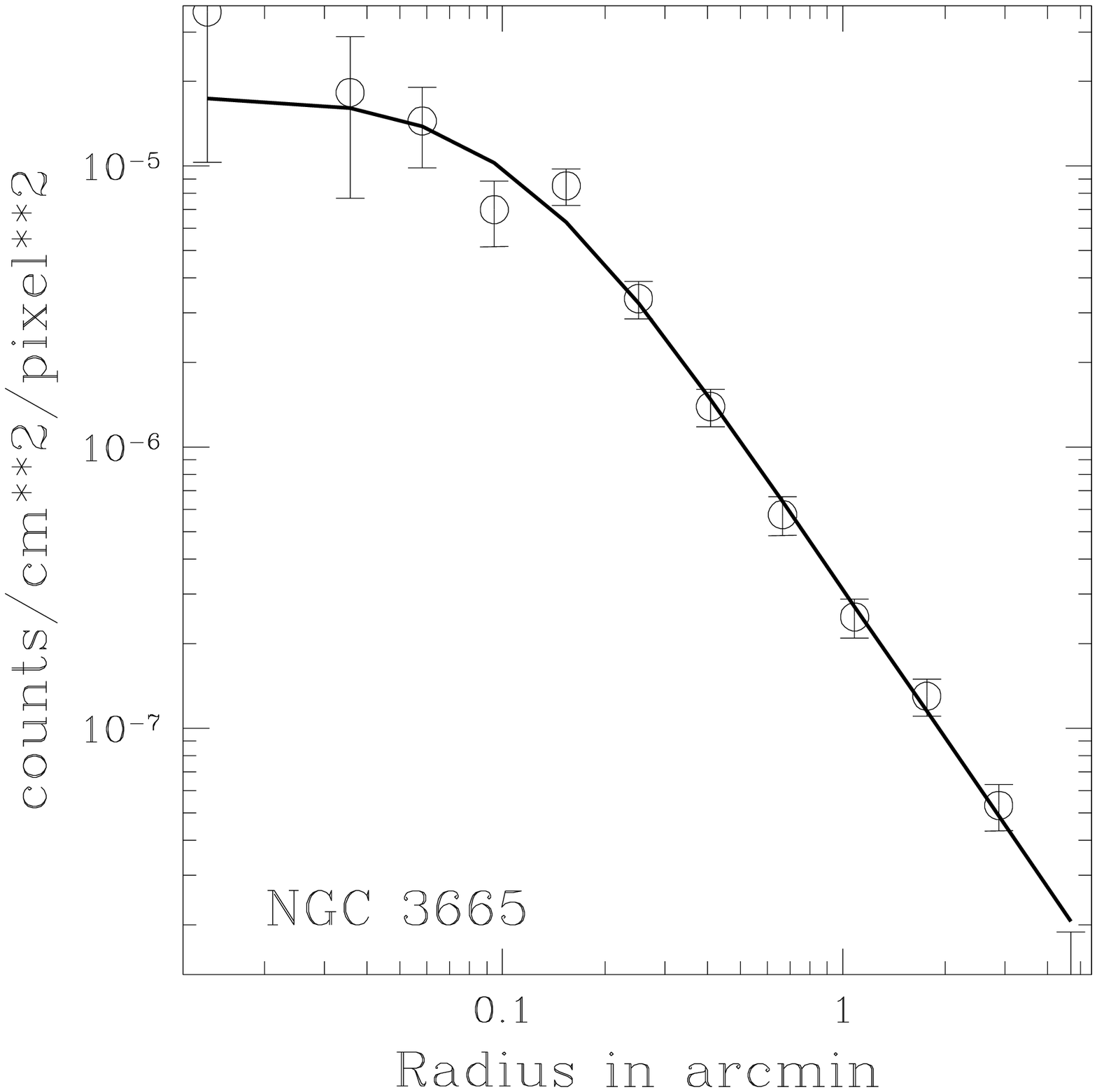}
\caption{\label{fig:spat1d}1D background subtracted surface brightness 
  profiles (projected from 2D) of NGC\,1587 (left) and NGC\,3665 (right). }
\end{figure}

The Cash statistic \citep{cash79} was used when minimising the 2D fits.
On its own this statistic does not give any indication of the quality of
fit, and it was originally intended to determine the quality of fit using a
Monte Carlo procedure similar to that used in \citet{helsdon00}. Using this
method, estimates of the quality of the fit were obtained by using the best
fit model to generate 1000 model datasets with added poisson noise. The
Cash statistic of each of these noisy datasets when compared with the best
fit model was then calculated, giving the expected distribution of the Cash
statistic for the best fit model. The real Cash statistic for the data was
then compared with this distribution, by for example, calculating the
number of standard deviations the real Cash value lay from the mean of the
1000 noisy models distribution. Unfortunately, it has now become apparent
that this procedure suffers from some serious flaws. Most notably, the
quality of fit obtained depends significantly on the binning of the data.

\subsubsection{Problems with testing fit quality}

To illustrate the problem with this method, consider the simple toy problem of 
a 256x256 image, with
20 counts in each pixel, apart from a 50 pixel wide bar running up the
image with 20.8 counts per pixel. This represents the correct model of the
data, from which can be generated 'observed' datasets by adding in poisson
noise. Given some 'observed' data, is it then possible to distinguish
between the datasets and a flat model with a value of $\sim$20.16 counts
per pixel (this level is equal to the average number of counts per pixel in
the original model)? The Monte Carlo method described above was used on the
256x256 image and also repeated with the data re-binned by a factor of 2, 4
and 8 (i.e. 128x128, 64x64, 32x32, with the overall counts in the image the
same in each case). For the 256x256 case the Monte Carlo method suggests
that the flat model is a perfectly reasonable fit to the data generated
with the bar model (the 'observed' Cash value lies 0.86$\sigma$ from the
distribution mean). For the 128x128 case, the flat model is found to be a
borderline acceptable fit (1.83$\sigma$), for 64x64 the flat model is ruled
out (3.43$\sigma$), and for the 32x32 case it is strongly ruled out (6.65
sigma offset). It should also be pointed out that this effect is not limited 
to the Cash statistic alone, as nearly identical results are obtained using 
$\chi^2$ statistics.

This shows that in the above case, the higher resolution data is more
likely to be considered an acceptable fit. The main reason for this is that
the width of the statistic distribution generated from the 1000 random
realizations increases roughly as the square root of the number of pixels,
while the size of the mismatch between the 'real' statistic value and the
distribution mean is approximately constant. This dependence on resolution
is clearly an undesirable property when trying to construct a reliable
estimate of the quality of fit.

The origin of this effect can be quite easily understood. When using either
the Cash statistic or the $\chi^2$ statistic, the value of the statistic
is derived from the addition of scaled residuals, without any account of
the \textit{spatial structure} of these deviations. For example, consider the
implications of structure in these deviations in the case of the $\chi^2$
statistic, where the maths is simpler. For a good model, data residuals
(data, $d_i$, minus model, $m_i$) are uncorrelated from one point to the 
next. This is clearly not the case in the above toy problem, as there is a
systematic offset across the image. Systematic offsets could also arise
very easily in surface brightness profile analyses. To see the effect of
systematic offsets on the $\chi^2$ statistic consider the following:

\begin{equation}
\chi^2 = \sum_{i} \frac{r_i^2}{d_i},
\end{equation}
where $r_i = d_i - m_i$. Now $r_i$ can be decomposed into a statistical 
contribution $e_i$, with an expectation value of zero, but with $<e_i^2>=m_i$, and a 
systematic misfit contribution $s_i$, which is due to real discrepancies
between the data and the model. 

\begin{equation}
\chi^2 = \sum \frac{e_i^2 + s_i^2 + 2e_is_i}{d_i}
\end{equation}

Now suppose that the number of bins is increased by a factor $f$, without
changing the total number of counts. Then the values of $d_i$ and $m_i$ drop
by the factor $f$. Taking expectation values throughout the above expression
for $\chi^2$, the sum runs over $fN$ values, $<e_i^2>$ and $d_i$ both scale 
as $1/f$, $s_i^2$ scales as $1/f^2$, and $<e_is_i>$ vanishes. This shows 
that the contribution to $\chi^2$ coming from the model misfit stays constant, 
whilst the contribution to $\chi^2$ from statistical variations, and the 
number of degrees of freedom, scale up as $f$. Since the width of the $\chi^2$ 
distribution scales as approximately the square root of the number of degrees 
of freedom, this means that the significance of the misfit will drop for the 
higher resolution data (or conversely, binning up the data is more likely 
to show a misfit), just as seen in the above example. The same effect 
also exists for the Cash statistic. It is also worth emphasising that this
resolution dependence only occurs if there is systematic structure in the
residuals. If the residuals do not show any systematic structure the $s_i^2$
term will scale as $1/f$ and the resolution dependence will disappear. e.g. if
in our simple toy model above we had distributed the pixels with 20.8 counts
randomly about the image, rather than grouped together in a bar, we would not
expect to see the fit quality get worse as the data are binned up (as is the
case when they are in a bar). Running the same test as was applied to the image
with a bar confirms this prediction -- in this case the fit is acceptable at all
resolutions (always $< 1\sigma$ from the distribution mean).

The resolution dependence of the fit statistic above is a demonstration
that simple misfit tests based on the value of the Cash or $\chi^2$ statistics 
do not take into account the effects of spatial correlations between residuals.
In particular, this means that in the limit of very fine binning, the misfit test
is a very weak measure of goodness of fit. As larger bins are used, spatially 
correlated residuals combine to become more significant, but ultimately all
structure will be lost in the limit of very coarse binning -- in the limit of a 
single bin, if the model and data have the same overall normalization, a 
perfect fit will always be obtained, regardless of the smaller scale structure
of the model and data. Given this, a reasonable compromise approach is to bin
on progressively larger scales, and to take the case where the fit is poorest
as an indication of the true quality of fit of the model. However, this is still
not ideal, since the scale of the systematic misfits could vary around the image,
so that no single binning scale is entirely satisfactory. It should also be noted
that these results are quite general, and apply, for example, to the use of $\chi^2$ 
to evaluate the adequacy of \textit{spectral} fits.

\begin{deluxetable}{lccc}
\tablecaption{\label{tab:spatfits1}Results of fitting a single beta 
model to the spatial emission.}
\tablewidth{0pt}
\tablehead{
\colhead{Group}    & \colhead{$\beta_{fit}$}          &  \colhead{Core radius (arcsec)}  & \colhead{Core radius (kpc)}     }
\startdata
NGC 1587 & 0.36$^{+0.01}_{-0.01}$ &  $<$1                   & $<$0.24                 \\
NGC 3665 & 0.46$^{+0.02}_{-0.02}$ &  6.1$^{+1.7}_{-1.6}$    & 0.83$^{+0.23}_{-0.22}$  \\
\enddata
\end{deluxetable}


The problems described above make it difficult to reliably estimate the
quality of our spatial fits. Running the 2D data for the two groups through 
the flawed test described above suggests that the circular beta models 
are acceptable fits. Running the quality of fit test at a variety of resolutions
(binning the data by factors of 1, 2, 4, 8 and 16)
suggests that there is systematic structure in the data minus model residuals 
of both groups (i.e. the quality of fit gets worse as the data are binned up). 
At the worse fit quality both groups are still acceptable fits although only
marginally so for NGC 1587 with a worse case offset of 1.99$\sigma$, while
NGC 3665 has a worse case offset of 0.5$\sigma$. However this result should
be treated with some care given the limits and reliability 
of the quality of fit test. It is however, reassuring to note that the 1D 
profiles (projected from 2D) shown in Figure~\ref{fig:spat1d} suggest that 
the models do appear to provide a reasonable match to the data.

\section{Comparison with previous ROSAT observations}
\label{sec:roscomp}

\begin{deluxetable}{lcccl}
\tabletypesize{\small} 
\tablecaption{\label{tab:rcomp1}Comparison of the X-ray spectral 
parameters derived in this \textit{Chandra} study with those from previous
\textit{ROSAT} studies.}
\tablewidth{0pt}
\tablehead{
\colhead{Instrument}          & \colhead{Temperature (keV)} & \colhead{Abundance (Solar)} & \colhead{log L$_X$ (erg s$^{-1}$)} & \colhead{Reference}}
\startdata
\cutinhead{\textbf{NGC 1587}}
\textit{Chandra}              &  0.37$^{+0.04}_{-0.04}$   & 0.012$^{+0.012}_{-0.009}$  & 41.22 $^{+0.05}_{-0.04}$ & 1 \\
\textit{Chandra}              &  0.53$^{+0.07}_{-0.04}$   & 0.3 (fixed)                & 40.88 $^{+0.01}_{-0.01}$ & 1 \\
\textit{ROSAT}                &  0.92 $\pm$ 0.15          & 0.3 (fixed)                & 41.10 $\pm$ 0.18         & 2 \\
\textit{ROSAT}                &  0.90$^{+0.81}_{-0.46}$$^*$   & 0.3 (fixed)                & 41.18 $^{+0.13}_{-0.41}$$^*$ & 3 \\
\cutinhead{\textbf{NGC 3665}}
\textit{Chandra}                &  0.45$^{+0.07}_{-0.04}$ & 0.04$^{+0.02}_{-0.01}$     & 40.58 $^{+0.05}_{-0.04}$ & 1 \\
\textit{Chandra}                &  0.52$^{+0.06}_{-0.04}$ & 0.3 (fixed)                & 40.37 $^{+0.01}_{-0.01}$ & 1 \\
\textit{ROSAT}                  &  0.45 $\pm$ 0.11        & 0.17 $\pm$ 0.14            & 40.81 $\pm$ 0.1          & 2 \\
\textit{ROSAT}                  &  0.33$^{+0.14}_{-0.07}$$^*$ & 0.3 (fixed)                & 40.77 $^{+0.37}_{-0.47}$$^*$ & 3 \\
\enddata
\tablecomments{\textit{ROSAT} Luminosities have been corrected to the exactly the same distances used in this work. 
Errors are 1 sigma unless marked with a $^*$ in which case they represent 90\% confidence limits.}
\tablerefs{(1) This Work, (2) \protect\citet{helsdon00}, (3) \protect\citet{mulchaey03}}
\end{deluxetable}

The spectral parameters as derived for these groups are compared with
previous ROSAT determinations in Table~\ref{tab:rcomp1}. The Chandra
derived parameters have a number of differences from the ROSAT values: The
temperature for NGC\,1587 is much lower than the ROSAT data had suggested,
while the derived bolometric luminosity is consistent with previous
estimates (although both are dependent on the abundance). In contrast the
X-ray temperature of NGC\,3665 is comparable to previous estimates and the
Chandra derived X-ray luminosity is a little lower. The ROSAT observations
were generally unable to put any strong constraints on the abundances, so
no meaningful comparison of the abundances is possible.

Given that there are clearly differences seen, it is important to try to
understand their origin. A likely reason is that the ROSAT data have been
contaminated by unresolved point sources which have been identified and
removed in the Chandra data. To check for this we first calculated the
total background subtracted counts in the Chandra data after removing all
identified contaminating sources. These counts were calculated in the ROSAT
bandpass (0.3--2.3keV) and over the same area that the ROSAT data were
derived (a 6 arcminute radius circle in both cases, with an additional
region excluded in NGC1587 where the ROSAT PSPC support ring fell in the
area). The contaminating source list from Chandra was then compared with
the equivalent list for the ROSAT data (from the analysis of
\citealt{helsdon00}), and the Chandra sources corresponding to sources
found in the ROSAT data were identified. The background subtracted counts
in the Chandra data were then re-calculated but this time, only the ROSAT
identified sources were removed. The results of these calculations for the
two groups are shown in Table~\ref{tab:rcomp2}.

\begin{deluxetable}{lccc}
\tabletypesize{\small}
\tablecaption{\label{tab:rcomp2}Net \textit{Chandra} counts in the 0.3-2.3 keV band, 
over approximately the same region the \textit{ROSAT} data were
extracted. The columns show firstly the net counts after removing all
contaminating sources as identified by the Chandra data, then the net
counts obtained if only sources also identified in the ROSAT data are
excluded. The final column shows the ratio between these two numbers,
and gives an indication of the fraction of the original \textit{ROSAT}
flux due to unresolved contaminating point sources.}
\tablewidth{0pt}
\tablehead{

\colhead{Group}              & \colhead{Chandra identified} & \colhead{ROSAT identified} & \colhead{Factor} \\
\colhead{}              & \colhead{sources removed} & \colhead{sources removed} & \colhead{difference}

}
\startdata
NGC 1587           &  616 $\pm$ 67        & 802 $\pm$ 68               & 1.30 \\
NGC 3665           &  666 $\pm$ 62        & 936 $\pm$ 64               & 1.41 \\
\enddata
\end{deluxetable}

Table~\ref{tab:rcomp2} shows that for both groups there is clearly
substantial contamination of the ROSAT data due to unresolved point
sources. For these groups unresolved point sources have increased the real
(Chandra measured) flux by 30--40\% in the ROSAT data. Taking into account
these unresolved point sources moves the ROSAT derived luminosities for
NGC\,3665 into better agreement with the Chandra data, whilst for NGC\,1587
the ROSAT luminosities drop relative to the Chandra luminosity. However the
luminosities can also be significantly altered due to differences in the
spectral properties (in particular the temperature). These luminosities are
based on unabsorbed fluxes, and the correction from an absorbed flux to
unabsorbed flux tends to be larger for cooler systems. For NGC\,1587 this
effect increases the luminosity by a factor of $\sim$1.4 (for a spectral
model with $T=0.9$ keV changing to a model with $T=0.37$ keV), which almost
exactly cancels out the effect of the unresolved point sources, resulting
in a similar final luminosity to that derived in the ROSAT data. Note that
the hotter temperature obtained when fixing the abundance at 0.3 results in
a luminosity lower than that obtained with the ROSAT data.

For NGC\,3665 the luminosity change due to the spectral model differences
is a factor of $\sim$1.2 which again acts to reduce the difference between
the ROSAT and Chandra data, with the end result that after allowing for
both this correction and the correction for unresolved point sources, that
the Chandra derived luminosity is a little below the ROSAT luminosity, but
probably just consistent within the errors. The difference in luminosity
between the Chandra data with free and fixed abundances, is again due to
the differences in the spectral models.

One final check that can be carried out on the spectral data is to examine
whether the unresolved point sources in the ROSAT data can explain the
spectral differences seen in the different instruments. To check this, the
integrated group spectra for each group were refit, but this time only
removing those Chandra sources also identified in the ROSAT data. For both
groups the effect of this was to raise the fitted temperature a little, to
$T=$0.46$^{+0.06}_{-0.09}$ for NGC\,1587 and $T=$0.49$^{+0.05}_{-0.04}$ for
NGC\,3665 (abundance free in both cases). With these sources added back in,
the luminosity for NGC\,3665 moves into very good agreement with the ROSAT
derived values, while the luminosity of NGC\,1587 drops a little due to the
change in spectral parameters.

In summary, adding back unresolved ROSAT sources into the Chandra data
results in good agreement between the temperature and luminosity for
NGC\,3665. However, for NGC\,1587, the temperatures are still lower than
the best fit ROSAT temperature, although this difference is reduced if a
fixed abundance of 0.3 is used, as with the ROSAT data.

\begin{deluxetable}{lcclcc}
\tabletypesize{\small}   
\tablecaption{\label{tab:rcomp3}spatial comparisons}
\tablewidth{0pt}
\tablehead{
\colhead{\textbf{NGC 1587}}        & \colhead{$\beta_{fit}$}           & \colhead{core radius} & \colhead{\textbf{NGC 3665}}        & \colhead{$\beta_{fit}$}           & \colhead{core radius} \\
                                   &                                   & \colhead{arcsec}      &                                    &                                   & \colhead{arcsec}      }
\startdata
\textit{Chandra}             &  0.36$^{+0.01}_{-0.01}$  & $<$1                 & \textit{Chandra}              & 0.46$^{+0.02}_{-0.02}$  & 6.1$^{+1.7}_{-1.6}$  \\
\textit{ROSAT} HP00           &  0.47 $\pm$ 0.06         & 20.4 $\pm$ 16.2      &\textit{ROSAT} HP00             &  0.49 $\pm$ 0.03        & 7.8 $\pm$ 7.2   \\
\textit{ROSAT} MDMB           &  0.42 $\pm$ 0.02         & $<$6                 &\textit{ROSAT} MDMB             &  0.42 $\pm$ 0.03        & $<$6            \\  
\enddata
\end{deluxetable}

Table~\ref{tab:rcomp3} compares the spatial parameters derived for the
Chandra data with those obtained in previous ROSAT analyses. As can be seen
the $\beta_{fit}$ values are roughly consistent with previous ROSAT values,
as are the core radii, although these are constrained much more tightly in
the higher resolution Chandra data. For NGC\,1587, the core radius is much
smaller than that obtained with ROSAT, although the ROSAT values are not
particularly well constrained, and the difference is not significant. The
slightly lower $\beta_{fit}$ value for NGC\,1587 is most likely related to
the smaller core radius as there is a well known correlation between core
radius and $\beta_{fit}$ in these models.

\section{Discussion}
\label{sec:dis}
Our primary result is that the bulk of the ``diffuse'' emission reported
from the ROSAT analyses of these two very low velocity dispersion groups is
confirmed by Chandra to be genuine hot plasma emission from a hot IGM.
Nonetheless, we find some significant differences between the ROSAT and
Chandra results for these two groups. These discrepancies appear to be
primarily related to the effects of contaminating point sources which are
not resolved in the ROSAT data. For these two groups these unresolved point
sources cause the ROSAT count rates to be overestimated by a factor of
1.3-1.4, and it should be reasonable to expect other low luminosity (log
$L_X \la$ 41.5) groups observed by ROSAT to show comparable contamination.
The remaining emission from these systems appears to mostly arise from a
hot gas component. The expected discrete source contribution from X-ray
binaries in the galaxies of these groups, is a factor of 7-10 times fainter
than the measured luminosities (assuming a discrete $L_X/L_B$ =
29.5 erg~s$^{-1} L_\odot^{-1}$ --- \citealt{osullivan01}), and would in any
case be restricted within the optical confines of the galaxies.

\begin{figure}[h]
\centering
\includegraphics[totalheight=10cm,angle=270]{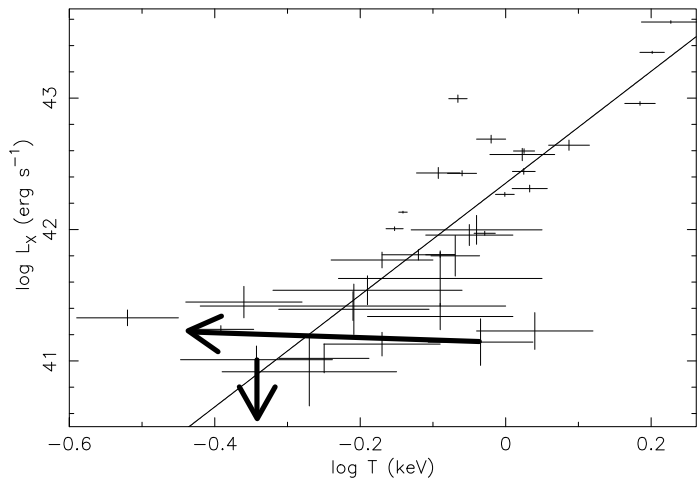}
\caption{\label{fig:lt}The $L_X:T$ relation from \protect\citet{helsdon00b} 
with arrows showing how NGC\,1587 and NGC\,3665 have moved as a result
of the analysis in this paper.}
\end{figure}

For a fixed spectral model, the removal of these point sources in the
Chandra data should result in lower group X-ray luminosities. However,
significant differences in the fitted spectral models (e.g. a large change
in temperature), may act to either increase or reduce the overall effect on
the inferred bolometric X-ray luminosity. In general, the properties we
derive for these two groups fall within the spread of other groups (e.g.
\citealt{mulchaey96,helsdon00,mulchaey03}), although they are amongst the
faintest and coolest groups currently known.For example, despite the large
change in temperature for NGC\,1587, and the drop in luminosity for
NGC\,3665, the groups still appear to be consistent with previously derived
scaling relations such as the $L_X:T$ relation, as shown in
Figure~\ref{fig:lt}. This suggests that their dynamical status is not
likely to be grossly different from that of other groups.

The low $\beta_{fit}$ values obtained are consistent with what has been
seen in other groups \citep{helsdon00,mulchaey03,osmond03}, and are broadly
consistent with what would be expected, given the possible trends of
$\beta_{fit}$ with temperature that have been reported (e.g.
\citealt{helsdon00,sanderson03}). As for the core radii of these two
systems, the Chandra data are able to put much stronger constraints on this
parameter than the ROSAT data. One of the problems in previous ROSAT
observations of groups (e.g. \citealt{helsdon00,mulchaey03}) is that often
the core radii have been unresolved. The Chandra data have enabled us to
derive tight constraints on the core radius of NGC\,3665, but, remarkably,
the core radius for NGC1587 is still unresolved in the Chandra data,
implying a power-law like surface brightness profile to within $\sim$ 0.5
kpc of the center of the system.

One might expect these results for beta and core radius ought to be fairly
reliable, since the data for these two systems are fit over a region 60 to
130 times larger than the implied core radius. However some caution is
needed, as it is possible that there may be a significant bias in these
results. In general, groups with good data quality require a 2-component
model to adequately describe their surface brightness profiles
\citep{mulchaey98,helsdon00}. Unfortunately, the Chandra data for NGC\,1587
and NGC\,3665 are not of sufficient quality to be able to constrain a
2-component model, or rule out the need for one. Where systems really are
described by a 2-component model, the use of a single component model can
result in a significant bias in the fitted parameters, in particular
$\beta_{fit}$ may be either under- or over-estimated \citep{helsdon00}.

A second potential problem with the issue of fitted surface brightness
profiles is the effect of aperture. Simulations of clusters generally
produce convex gas density profiles, and as a result the $\beta_{fit}$
value derived for a surface brightness profile may depend strongly on the
range of radii used in the fit \citep{navarro95,bartelmann96,voit02}. In
particular, $\beta_{fit}$ values derived on scales much less than the
virial radius tend to be systematically low. This could be a significant
problem for both our groups, as they are both detected to only
$\sim$0.1-0.2 of their virial radius ($R_{200}$). However, such simulations
are by no means guaranteed to include the correct gas physics, so one would
like to appeal directly to observations on the question of whether gas
density profiles follow a power law at large radii. Recent XMM observations
of two richer galaxy groups \citep{rasmussen03}, in which X-ray emission is
traced to $\sim$0.65$R_{200}$, are encouraging, in that a beta model is
found to provide a good representation of the profiles to the edge of the
data, with no significant convexity in the measured profiles.

Assuming that our surface brightness profiles do provide a reasonable
description of the data, it is possible to derive a number of other
interesting properties for these systems, such as gas mass and total mass.
In fact, within the data extraction radius, the gas mass is relatively
insensitive to changes in $\beta_{fit}$ and core radius, as the
normalisation of the gas density profile is constrained by the
normalisation of the spectral fit. However, when making large
extrapolations of the data, the masses can be quite sensitive to changes in
$\beta_{fit}$ and the results from large extrapolations should be viewed
with some caution.

\begin{figure}[t]
\centering
\includegraphics[totalheight=10cm,angle=270]{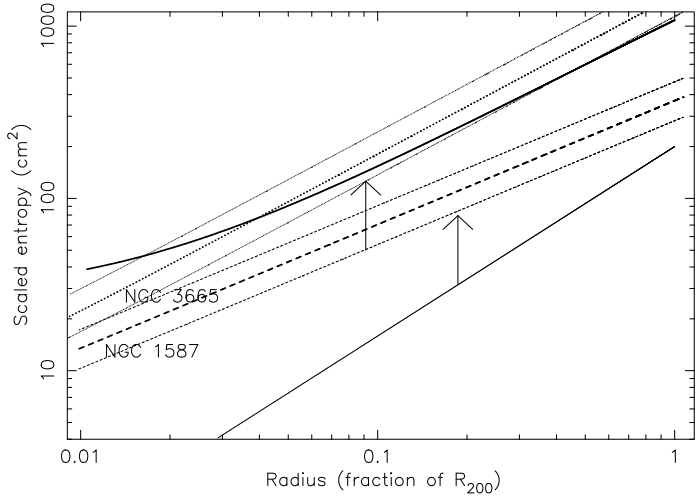}
\caption{\label{fig:entropy}Scaled entropy (S/T) as a function of 
  $R_{200}$ for NGC\,1587 (dashed line) and NGC\,3665 (dotted line). The
  bold solid line is the average scaled entropy profile for groups as
  determined by \protect\citet{ponman03}. The fainter solid line (lower
  one) shows the expected canonical relationship, normalized to a 0.4 keV
  system (see text for details). The two arrows mark the extent of the data
  for the two groups in this analysis. The faint dashed and dotted lines show
  the approximate 1 sigma error bounds for the two profiles.}
\end{figure}

Given spectral information, together with a model for the surface
brightness distribution, it is possible to derive the gas entropy in these
systems. Entropy profiles were derived by assuming an isothermal
temperature distribution in each, and inferring the gas density profile
from a combination of the spectral data and the surface brightness profile.
The entropy, defined here as $S=T/n_e^{2/3}$, where $n_e$ is the electron
density, is then readily calculated. For self-similar systems virialising
at the same epoch (and hence having the same mean overdensity) one expects
the entropy at a given overdensity radius to simply scale with mean system
temperature \citep{ponman99}. Hence, for comparison between different
systems, it is helpful to scale entropies by $1/T$. In
Figure~\ref{fig:entropy} we show such scaled entropy profiles for NGC\,1587
and NGC\,3665.  Also shown for comparison is the average scaled entropy
profile for a number of low temperature systems, derived by
\citet{ponman03}, and a line representing expectations from simulations
involving only gravitational physics and shock heating. The latter follows
the trend $S\propto r^{1.1}$ expected from analytical theories of spherical
accretion of gas onto clusters \citep{tozzi01}, and found in cosmological
simulations. This profile has been normalised using the simulations of
\citet{voit03b}, and scaled to $T=0.4$~keV, to match the typical
temperatures of NGC\,1587 and NGC\,3665.  This shows that both groups have
entropies well above the self-similar expectations, as is typically the
case in low mass systems \citep{ponman03,voit03}. The scaled entropy of the
IGM in NGC\,3665 agrees well with that seen in other poor groups, whilst
that in NGC\,1587 is unusually low, accounting for the location of the
latter to the high side of the $L_X:T$ relation (Fig.\ref{fig:lt}).  It has
already been pointed out by \citet{mushotzky03} and \citet{sun03} that the
scaled entropy in groups shows considerable real scatter, which may reflect
differences in the star-forming and merger histories within different
groups.

\begin{deluxetable}{lcccccc}
\tabletypesize{\scriptsize} 
\setlength{\tabcolsep}{0.02in}
\tablecaption{\label{tab:mass}Group masses and related properties as determined under the assumption of an isothermal model.}
\tablewidth{0pt}
\tablehead{

\colhead{}                    & \multicolumn{3}{c}{\textbf{NGC 1587}}                                                                               &  \multicolumn{3}{c}{\textbf{NGC 3665}} \\
\colhead{}                    & \colhead{R$_{extract}$}               & \colhead{0.3R$_{200}$}               & \colhead{R$_{200}$}                  & \colhead{R$_{extract}$}              & \colhead{0.3R$_{200}$}               & \colhead{R$_{200}$} }
\startdata
Total Mass ($h_{75}^{-1}$M$_\sun$)   & 9.04$^{+0.98}_{-0.98}\times10^{11}$  & 1.39$^{+0.15}_{-0.15}\times10^{12}$ & 4.58$^{+0.50}_{-0.50}\times10^{12}$ & 1.04$^{+0.16}_{-0.10}\times10^{12}$ & 2.70$^{+0.42}_{-0.24}\times10^{12}$ & 8.97$^{+1.39}_{-0.78}\times10^{12}$ \\
Gas Mass ($h_{75}^{-2.5}$M$_\sun$)   & 2.20$^{+0.57}_{-0.44}\times10^{10}$  & 5.00$^{+1.29}_{-1.00}\times10^{10}$ & 5.01$^{+1.29}_{-1.01}\times10^{11}$ & 4.35$^{+1.38}_{-0.88}\times10^{9}$  & 2.03$^{+0.64}_{-0.41}\times10^{10}$ & 1.42$^{+0.45}_{-0.29}\times10^{11}$ \\
Stellar Mass ($h_{75}^{-2}$M$_\sun$) & $\sim2.5\times10^{11}$               & $\sim2.5\times10^{11}$              & $\sim3.1\times10^{11}$              & $\sim1.2\times10^{11}$              & $\sim1.5\times10^{11}$              & $\sim1.6\times10^{11}$           \\
Stellar Mass Fraction ($h_{75}^{-1}$)& $\sim0.28$                           & $\sim0.18$                          & $\sim0.06$                          & $\sim0.12$                          & $\sim0.06$                          & $\sim0.02$              \\
Gas Mass Fraction($h_{75}^{-1.5}$)   & 0.024$^{+0.010}_{-0.007}$            & 0.036$^{+0.015}_{-0.010}$           & 0.109$^{+0.045}_{-0.031}$           & 0.0042$^{+0.0019}_{-0.0013}$        & 0.0075$^{+0.0033}_{-0.0023}$        & 0.016$^{+0.007}_{-0.005}$         \\
Baryon Fraction                      & $\sim0.3$                            & $\sim0.22$                          & $\sim0.17$                          & $\sim0.12$                          & $\sim0.07$                          & $\sim0.04$              \\
Star formation efficiency            & $\sim0.92$                           & $\sim0.83$                          & $\sim0.38$                          & $\sim0.97$                          & $\sim0.88$                          & $\sim0.53$              \\
Mass-to-light ratio ($h_{75}^{1}$M$_\sun$/$L_B$) & $\sim22$                      & $\sim34$                            & $\sim58$                            & $\sim44$                            & $\sim89$                            & $\sim278$               \\
\enddata
\end{deluxetable}

In Table~\ref{tab:mass} we show a number of inferred parameters for our two
groups, assuming spherical symmetry and an isothermal IGM. The errors on
the parameters are obtained by taking the extreme effects of the errors on
the spectral normalisation and temperature. For each group we show the
parameters calculated at 3 different radii: the data extraction radius,
0.3$R_{200}$ and $R_{200}$. $R_{200}$ is calculated by deriving the
over-density profile, and finding the radius which corresponds to an
over-density of 200 (relative to the critical density of the Universe).
For NGC\,1587 this radius was 327 kpc (factor of 5 extrapolation in radius)
and for NGC\,3665, 409 kpc (factor of 9 extrapolation in radius).

As can be seen, these systems have fairly low total masses -- the typical
mass of the groups given in \citet{mulchaey96} is $\sim 1\times10^{13}$
M$_\sun$, although those masses are derived to a variety of different
radii. Our masses and group temperatures are roughly consistent with the
isothermal mass-temperature relation for groups and clusters derived by
\citet{sanderson03}. The gas masses are fairly low for groups, although due
to the flat gas density profiles inferred, they rise strongly with radius,
as do the gas mass fractions. Despite the comparatively low gas mass and
gas fractions, these values are still substantially larger than those of
bright ellipticals (e.g. \citealt{bregman92,osullivan03}) at 0.3$R_{200}$,
and the values are rising more rapidly with radius in these groups, since
\citet{osullivan03} find that typically $\beta_{fit}\sim 0.5$ in the halos
of early-type galaxies.

In addition to the total and gas masses we also show a number of other
parameters. The stellar mass is estimated by searching NED for galaxies
within the appropriate radius and assuming mass-to-light ratios of
5~$M_{\sun}/L_B$ for early type galaxies and 1~$M_{\sun}/L_B$ for late
types. Both these systems are very poor, and most of the brighter member
galaxies are found in the central regions, leading to stellar mass fraction
which is high in the central region, but drops off fairly rapidly with
radius. The large stellar mass fraction also leads to a high mean star
formation efficiency (defined as $M_*/(M_*+M_{gas}$) in the central
regions. Clusters typically have efficiencies of $\sim$ 0.2-0.3
\citep{david90,arnaud92}, however once again these are generally derived at
a much larger fraction of R$_{200}$ than the radius to which groups are
detected, and the extrapolation to R$_{200}$ for the groups shows that
there is the potential for a substantial drop with radius.

The baryon fractions and mass-to-light ratios are both roughly consistent
with the range of values seen in \citet{mulchaey96}. The baryon fraction
shows less radial variation than most of the other parameters, as the
effect of the rising gas mass fraction and dropping stellar mass fraction
partially cancel out to produce a flatter baryon fraction profile. The
baryon fraction of NGC\,1587 is roughly consistent with the typical values
of 0.1-0.3 seen in clusters (e.g. \citealt{david95,hradecky00}), while
NGC\,3665 is a little lower than this.

Thus overall these two systems have properties which appear to be broadly
consistent the scaling relations observed in other galaxy systems (apart
from the $L:\sigma$ relation as discussed below). In the few cases where
there may be some significant differences (e.g. star formation efficiency)
these two groups tend to differ in the sense that they appear more like
individual galaxies, although some care is needed as some of these
properties have the potential for significant radial changes. For example,
the star formation efficiency is clearly dropping with radius, and at
values close to R$_{200}$ this value could be consistent with clusters
given the uncertainties involved in the extrapolation.

We conclude that in most respects, NGC\,1587 and NGC\,3665 appear to be
fairly normal, low mass, groups. Given this, there is no reason to expect
them to have a significant offset from the group $L:\sigma$ relation.
However as discussed in the introduction, these two groups are
significantly offset from an extrapolation of the cluster $L:\sigma$
relation, and this offset is in a direction opposite to what would be
expected, given the effects of preheating. If the X-ray luminosities really
are comparatively normal, especially in the case of NGC\,3665, then perhaps
the velocity dispersions are unusual in some way.

As mentioned earlier in \S\ref{sec:reduction}, the original velocity
dispersions of these systems were estimated from very few galaxies, which
suggests that they may be suspect. However, as shown above, the addition of
related galaxies from NED was able to significantly increase the group
membership, and confirm the low velocity dispersions. One can attempt to
further refine these dispersions by only including galaxies within the
virial radius in projection (1 Mpc was used earlier). This has the effect
of dropping the membership to 6 galaxies for both groups, leading to a
velocity dispersion of 101 $\pm$ 29 for NGC\,1587 and 61 $\pm$ 17 for
NGC\,3665 -- values which are actually almost identical to those obtained
under the less strict membership requirements used in
\S\ref{sec:reduction}.

\begin{figure}[t]
\centering
\includegraphics[totalheight=10cm,angle=270]{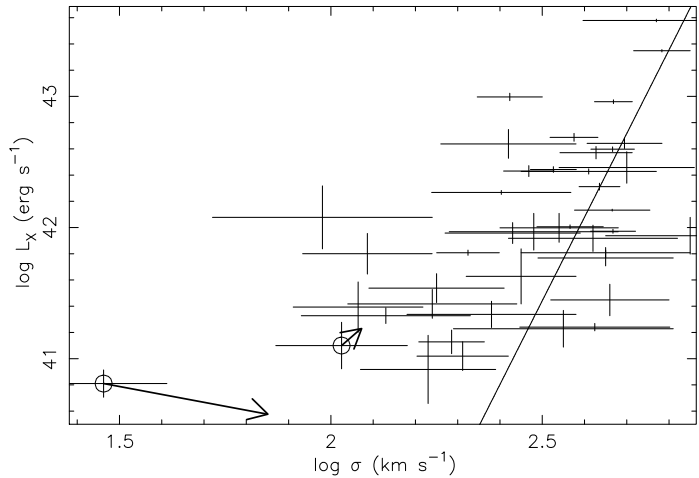}
\caption{\label{fig:lvplot}The $L_X:\sigma$ relation using data from 
  \protect\citet{helsdon00b} with arrows showing how NGC\,1587 and
  NGC\,3665 have moved as a result of the analysis in this paper. The line
  marks the cluster line as derived by \protect\citet{white97}.}.
\end{figure}

The measured velocity dispersion for NGC\,3665 is immediately suspect, as a
virialised system must have a minimum mean density related to the density
of the universe, which can be used to constrain the velocity dispersion of
a virialised group to be at least 100-200~km~s$^{-1}$ \citep{mamon94}. The
velocity dispersion of NGC\,3665 is clearly below this limit and the
dispersion of NGC\,1587 is also uncomfortably low. Even after correcting
for statistical biases in the group velocity dispersion (the standard
method of calculating velocity dispersions is biased low, and this bias
is most significant when the number of group members is low -- Helsdon
\& Ponman in prep.) these velocities are only increased to 117 km s$^{-1}$ and 70 km s$^{-1}$
respectively.  Figure~\ref{fig:lvplot} shows how NGC\,1587 and NGC\,3665
move on the group $L:\sigma$ relation if these velocity dispersions are
used along with the Chandra luminosities derived earlier. The other data
points in Figure~\ref{fig:lvplot} are taken from \citet{helsdon00b}.

Since we have mass profiles from our X-ray analysis, we can predict the
velocity dispersion we would expect to see within the group, under given
assumptions about the galaxy orbits. In the simple case of an isothermal
sphere, with isotropic velocities, $\sigma^{2}(r)=G M(r)/2r$, and using the
masses given in Table~\ref{tab:mass}, we obtain a predicted velocity
dispersion for NGC\,1587 of $\sim$ 174 km s$^{-1}$, and for NGC\,3665 of
$\sim$ 217 km s$^{-1}$ (these predictions are essentially the same for each
of the three mass/radius combinations in the table) -- both much larger
than the measured values.  \citet{lokas01} calculated the line-of-sight
velocity dispersions which would be observed for galaxies orbiting within
halos with density profiles which follow the ``NFW'' profile of
\citet{navarro95}, for a range of halo concentration parameters and orbital
anisotropies. In general, these results show that the isothermal sphere
assumption gives a reasonable approximation to the integrated value of
$\sigma$ for likely halo concentrations.  Purely radial orbits result in
$\sigma(r)$ profiles which rise sharply in the centre, whilst circular
orbits result in a central minimum in $\sigma(r)$. However, for velocity
dispersions calculated (as here) for galaxies falling within $R_{200}$, all
reasonable models produce mean velocity dispersions which fall within $\sim
20$\% of the isothermal value.

If the velocity dispersion values predicted from the masses above are a
fair estimate of the `real' velocity dispersion, then some process must
have reduced the velocity dispersion of the galaxies in these systems. One
possible candidate is dynamical friction. The timescale for dynamical friction
in these two groups can be estimated using the equations given by 
\citet{binney87} which integrate the \citet{chandrasekhar43} dynamical
friction force as a galaxy falls towards the group centre. We assume that
the total gravitational mass is proportional to $r^{-2}$, and that a galaxy
is initally on a circular orbit at a radius of 200 kpc (a little below the
predicted virial radius for NGC 3665, and approximately the average radius
of galaxies in NGC 1587) with a velocity of $\sqrt{2} \sigma_{p}$, where 
$\sigma_{p}$ is the predicted velocity dispersions for the group as described
above. Under these assumptions an $L_*$ galaxy with a typical mass to light
ratio of 10 $M_{\sun}/L_{\sun}$ is predicted to fall to the group center in
$\sim 2.9$ Gyr for NGC 1587 and $\sim 3.9$ Gyr for NGC 3665. Thus the 
effects of dynamical friction acting over a substantial fraction of the
Hubble time could be significant in both these groups.

Dynamical friction leads to a transfer of energy from a large orbiting
body, to the sea of dark matter particles through which it moves.  Another
possibility is that orbital energy may be converted into internal energy of
galaxies, via tidal interactions. This effect is not significant in
clusters, since the orbital velocities of galaxies are substantially higher
than their internal velocity dispersions. In the present case, however, the
reverse is true, and strong tidal encounters between the slowly moving
galaxies could absorb substantial orbital energy, leading to
circularisation and lowering of orbits. Both effects would reduce measured
line-of-sight (los) velocity dispersions to some extent.

A third possibility, is that in NGC\,1587 and NGC\,3665, much of the
orbital motion happens to take place in the plane of the sky, and does not
contribute to the line of sight velocity dispersion. Since groups generally
form within cosmic filaments, in cosmological simulations, it would not be
surprising to find significant elongation and an anisotropic velocity
dispersion tensor in many systems (e.g. \citealt{tovmassian02}).
Distributing such systems at a variety of angles to the los, will then
result in additional scatter (over and above that expected from random
sampling of an isotropic velocity distribution) in $\sigma$. Such
non-statistical scatter is certainly observed in the $\sigma:T$ relation
\citep{helsdon00b}.  In this picture, low $\sigma$ groups have
3-dimensional velocity dispersions which are substantially more than
$\sqrt{3}\sigma_{los}$, and there will be other groups for which the
reverse is true.

\section{Acknowledgments}

The authors would like to thank Ben Maughan for helpful discussions on the
Cash statistic and for assistance with the Cash statistic simulations. We
would also like to thank Gary Mamon for helpful discussion of the velocity
dispersion in groups. Partial support for this project was provided by SAO
grants GO1-2134X and AR3-4013X.

\bibliography{apj-jour,../../bibtex/reffile} 
\label{lastpage} \end{document}